\documentclass{iacrtrans}

\usepackage{braket}
\usepackage{float}
\usepackage{tikz}
\usepackage{algorithm}  
\usepackage{algpseudocode}  
\usepackage{amsmath}  
\usepackage{cite}
\usepackage{makecell}
\author{ Anpeng Zhang, Xiutao Feng}
\institute{
   Academy of Mathmatics and Systems Science Chinese Academy of Science, Beijing, China, \email{{zhanganpeng, fengxt}@amss.ac.cn}
}

\title{Applications of Quantum Annealing in Cryptography}

\begin{document}

\maketitle


\begin{abstract}
This paper presents a new method to reduce the optimization of a pseudo-Boolean function to QUBO problem which can be solved by quantum annealer. The new method has two aspects, one is coefficient optimization and the other is variable optimization. The former is an improvement on the existing algorithm in a special case. The latter is realized by means of the maximal independent point set in graph theory. We apply this new method in integer factorization on quantum annealers and achieve the largest integer factorization(4137131) with 93 variables, the range of coefficients is [-1024,1024] which is much smaller than the previous results. We also focus on the quantum attacks on block ciphers and present an efficient method with smaller coefficients to transform Boolean equation systems into QUBO problems.
\end{abstract}

\section{Introduction}
With the rapid development of quantum computing technology, quantum computer has gradually become a reality. In fact, noisy intermediate-scale quantum (NISQ)\cite{preskill2018quantum} computers with 10-80 qubits have been made by some laboratories such as IBM and Google\cite{nay2019ibm}. Unfortunately, these quantum computers on this scale have only theoretical value. For example, today we can decompose $15=5 \times 3$ with Google's quantum computer\cite{vandersypen2001experimental}, but we can't use it to attack RSA, which has been shown to be unsafe under quantum computers\cite{shor1999polynomial}. So this paper focuses on another kind of quantum computers on a much larger scale---quantum annealers.

Quantum annealing processors naturally return low-energy solutions\cite{kadowaki1998quantum}\cite{messiah2014quantum}; some applications require the real minimum energy (optimization problems)\cite{humble2014integrated} and others require good low-energy samples (probabilistic sampling problems)\cite{negre2020detecting}. In fact, compared with traditional quantum computers, quantum annealers can only handle with specific optimization problem, which is called QUBO(Quadratic Unconstrained Binary Optimization) problem\cite{patton2019efficiently}. 

In mathmatics, QUBO is also called the optimization of quadratic pseudo-Boolean functions\cite{hammer1969pseudo}. This problem has wide applications in both mathematics and cryptography. In 2002, Endre Boros and Peter L. Hammer\cite{boros2002pseudo} introduced pseudo-Boolean functions when they studied the weighted maximal independent vertex sets of graphs and presented an algorithm to reduce higher order pseudo-Boolean functions to quadratic. They also introduced a new way to express pseudo-Boolean functions which is called $posifrom$. Further more, they established a one-to-one correspondence between the weighted maximal independent vertex set of a graph and the maximum value of a pseudo-Boolean function. 

In cryptography, integer decomposition has always been one of the basic problems of public-key cipher\cite{rivest1978method}. In 2018, Jiang $et \ al$\cite{jiang2018quantum} introduced the above method into integer decomposition for the first time, and creatively replaced the original equation with the multiplication table, which can greatly reduce the equation coefficients. They successfully realized the decomposition of $376289$ on D-Wave quantum annealer. Wang $et \ al$ \cite{peng2019factoring}\cite{wang2020prime}improved Jiang's method and increased the number to $1028171$. 

As for symmetric cipher, in 2021, Burek $et \ al$\cite{burek2022algebraic} presented an algebraic attack on block ciphers using quantum annealing. They designed an algorithm to transform algebraic equations of symmetric cipher into the QUBO problems and applied their algorithm to AES-128. Unfortunately, the number and coefficients of variables seem too large to run on existing quantum annealing machines.
\subsection{Our contribution}
In this paper, we improve Boros' algorithm about reducing higher order pseudo-Boolean functions to quadratic and present a new algorithm which can greatly reduce the range of coefficients(\textbf{Algorithm 2}). We introduce the concept of posiform in graph theory to optimize the number of variables of QUBO problems for the first time and apply our method in integer decomposition and algebraic attack on block ciphers. As a result, we decompose $4137131$ with $93$ variables and the range of coefficients is $[-1024,1024]$ which is much smaller than the previous methods. In algebraic attack, we give a general method to transform Boolean equations into QUBO, and estimate the upper bounds of the number of variables and coefficient range. The two upper bounds both are polynomial functions about the number of variables and the coefficients of the original equations.

\section{Preliminaries}
\subsection{Posiform and the stability of graphs}
A Mapping $f$: $F_2^n \to R$ is called a $pseudo$-$Boolean$ $function$. All pseudo-Boolean functions can be uniquely represented as $multi$-$linear$ $polynomials$, of the form
\[ f(x_1,...,x_n)=\sum_{S\subseteq [n]}c_S\prod_{j\in S}x_j ,\]
where $[n]=\{1,...,n\}$, by convention, we shall always assume that $\prod_{j\in \emptyset}x_j=1$.
Pseudo-Boolean functions are also represented as $posiforms$, i.e. polynomial expressions in terms of all the literals, of the form
\[ \phi(x_1,...,x_n)=\sum_{T\subseteq L}a_T\prod_{u\in T}u ,\]
where L=$\{x_1,\overline{x_1},...,x_n,\overline{x_n}\}$ denote the set of literals, $a_T \ge 0$ whenever $T \neq \emptyset$.
Since $u\overline{u}=0$ holds for all $u \in F_2$, it is customary to assume that $a_T=0$ if $\{u,\overline{u}\}\subseteq T$ for some $u\in L$. 

Given a graph $G=(V,E)$, its $stability$ $number$ $\alpha (G)$ is defined as the maximum size of its all maximal independent vertex sets. Furthermore, if there is a weight $\omega: V \rightarrow R_{+}$ associated to the vertices, then the $weighted$ $stability$ $number$ $\alpha_{\omega}(G)$ of $G$ is defined as the maximum weight of its maximal indenpendent vertex sets.

Given a posiform $\phi$, we then associate $\phi$ with a weighted graph $G_{\phi}(V,E)$, called $conflict$ $graph$ constructed as follows: $V=\{T\in L | T\neq \emptyset, a_T \neq 0\}$, $E=\{(T, T')| T, T' \in E, \exists u \in T \ s.t. \ \overline{u} \in T' \}$. To a vertex $v \in E$ we shall associate $a_T$ as its weight. The $weighted$ $stability$ $number$ of $G_{\phi}$ is written as $\alpha_{a}(G_{\phi})$.

Given a graph or a posiform, the following two interesting connections are shown\cite{boros2002pseudo}:
\\
\\
\textbf{Theorem 1.} $For\ any\ posiform\ \phi,$
$$max_{x\in {F_2^n}}\phi(x)=a_{\emptyset}+\alpha_{a}(G_{\phi}).$$
\\
\\
\textbf{Theorem 2.} \textit{Given a graph $G(V,E)$ and nonnegative weights $a_i \ge 0$ associated to the vertices $i\in V$, there exists a posiform $\phi_{G}$ in $n' < |V|$ variables, consisting of $|V|$ terms, and such that}
$$\alpha_{a}(G)= max_{x\in {F_2^{n'}}}\phi_{G}(x).$$
\subsection{Quantum annealing}
In quantum physics, a quantum fluctuation (also known as a vacuum state fluctuation or vacuum fluctuation) is the temporary random change in the amount of energy in a point in space, as prescribed by Werner Heisenberg's uncertainty principle. Quantum annealing (QA) is an optimization process for finding the global minimum of a given objective function over a given set of candidate solutions (candidate states), by a process using quantum fluctuations. Quantum adiabatic computation (QAC), as developed by Farhi et al., approaches the same task given a complex Hamiltonian whose ground state
encodes the solution to the optimization problem. This computation begins in the ground state of a simple, well-characterized Hamiltonian, which is then adiabatically evolved to the complex, problem Hamiltonian. According to
the adiabatic theorem, the system state will also evolve the ground state of the problem Hamiltonian provided the
evolution is sufficiently slow to prevent excitations to any higher-lying state. At the end of the annealing process, the
measured qubits will encode the optimal solution to the problem within a bounded degree of certainty (due to noise
within the closed system).

The time-dependent Hamiltonian of the quantum system is given by combining the initial Hamiltonian and the final
Hamiltonian\cite{albash2018adiabatic}.
$$H(t)=(1-\frac{t}{T})H_B+\frac{t}{T}H_P.$$
Here $H_B$ is the initial Hamiltonian and $H_P$ is the final Hamiltonian written as follows.
\[H_B=-\sum{\delta_{x}^{(i)}} \]
\[H_P=\sum{h_i\delta_z^{(i)}}+\sum{J_{ij}\delta_z^{(i)}\otimes\delta_z^{(j)}}\]
Where Pauli operator $\delta_x$ defines the $x$-basis, $\delta_z$ defines the $z$-basis, written as $\begin{pmatrix} 1&0\\0&{-1}\end{pmatrix}$ . The eigenvalue of $H_P$ gives the total energy of the system. 

The time-dependent Hamiltonian $H(t)$ of the physical system evolves according to Schrödinger equation
$$i\frac{d}{dt}\ket{\psi(t)} =H(t)\ket{\psi(t)}$$
where $\ket{\psi(t)}$ is the state of the system at any time $t\in[0, T]$. Let $\ket{\phi_i(t)}$ be the $i$-th instantaneous eigenstate of $H(t)$,
that is, $H(t)\ket{\phi_i(t)} = E_i(t)\ket{\phi_i(t)}$ holds through the entire evolution. If the system is initialized in the ground state
$\ket{\phi_0(t=0)}$, then the evolution proceeds slow enough to avoid exciting to the higher-lying eigenstates, e.g.,$\ket{\phi_1(t)}$.
Ultimately, the system will be prepared in the instantaneous ground eigenstate $\ket{\phi_0(t=T)}$.

Consider the eigenvalues of $H_P$, denote the eigenvector of $\delta_z^{(i)}$ as $\bf{x_{a_i}^{(i)}}$ ,where $a_i$ is the eigenvalue corresponding to the eigenvector, $a_i\in \{1,-1\}$. One can easily check that 
\[ H_P(\otimes{\bf{x_{a_i}^{(i)}}})= (\sum{h_ia_i}+\sum{J_{ij}a_ia_j})(\otimes{\bf{x_{a_i}^{(i)}}}).\]
So $\otimes{\bf{x_{a_i}^{(i)}}}$ is an eigenvector of $H_P$ with the eigenvalue $\sum{h_ia_i}+\sum{J_{ij}a_ia_j}$. Since $\otimes{\bf{x_{a_i}^{(i)}}}$ can run all $2^n$ eigenvectors of $H_P$, all eigenvalues of $H_P$ have the form $\sum{h_ia_i}+\sum{J_{ij}a_ia_j}$. The minimum energy of the system is the minimum of the function $\sum{h_ia_i}+\sum{J_{ij}a_ia_j}$.
\section{QUBO problem}
QUBO(Quadratic Unconstrained Binary Optimization) is a significant problem with many applications in the field of computing, in this section we show how to transform a high-order pseudo Boolean function optimization problem into QUBO.
\subsection{Reductions to quadratic optimization}
\textbf{Observation 1.} Assume that $x, y, z \in F_2$. The following equivalences hold:
\begin{equation}
\begin{split}
      xy&=z  \  \ iff \  \ xy-2xz-2yz+3z=0 \\
      xy&\neq z \ \ iff \ \ xy-2xz-2yz+3z>0.
\end{split}
\end{equation}

Then we can reduce the optimization of a pseudo-Boolean function to the optimization of a quadratic pseudo-Boolean function by the folllowing algorithm.
\begin{algorithm}[H]
  \caption{ReduceMin($f$)}  
\hspace*{0.02in} {\bf Input:} 
A pseudo-Boolean function $f$ given by its multi-linear polynomial form.\\
\hspace*{0.02in} {\bf Output:} 
A quadratic pseudo-Boolean function $g$.
 \begin{algorithmic}[1] 
\State {\bf Initialize:} Set $M=1+2\sum_{S\subseteq [n]}|c_S|$, $m=n$, and $f^n=f$.
\While{there exists a subset $S^*\subseteq[n]$ for which $|S^*|>2$ and $c_{S^*}\neq 0$}
\State 1. Choose two elements $i$ and $j$ from $S^*$ and update
\State$c_{\{i,j\}}=c_{\{i,j\}}+M$, set
\State$c_{\{i,m+1\}}=c_{\{j,m+1\}}= -2M$ and
\State$c_{\{m+1\}}= 3M$, and
\State for all subset $\{i,j\}\subseteq S$ with $c_S \neq 0$ define 
\State  $c_{(S\backslash \{i,j\})\cup \{m+1\}}=c_S$ and set $c_S=0$.
\State 2. Define $f^{m+1}(x_1,...,x_{m+1})=\sum_{S\subseteq [n]}c_S\prod_{k\in S}{x_k}$, and set $m=m+1$. 
\EndWhile
\State Output: $g=f^m$
    \label{code:recentEnd}  
  \end{algorithmic}  
\end{algorithm}

This algorithm was proposed by Boros in 2002. We generalize his algorithm based on the following observation.
\\
\textbf{Observation 2.} Assume that $x_1, x_2,...,x_n, x_{n+1} \in F_2$. The following equivalences hold:
\begin{equation}
\begin{split}
      x_1x_2...x_n&=x_{n+1}  \  \ iff \  \ x_1x_2...x_n-2\sum_{i=1}^{n}x_ix_{n+1}+(2n-1)x_{n+1}=0 \\
       x_1x_2...x_n&\neq x_{n+1}  \  \ iff \  \ x_1x_2...x_n-2\sum_{i=1}^{n}x_ix_{n+1}+(2n-1)x_{n+1}>0. 
\end{split}
\end{equation}
\textbf{Proof.} There are three cases of the relationships between $x_1x_2...x_n$ and $x_{n+1}$.
\\
$Case1$:  $x_1x_2...x_n-x_{n+1}=1$, then we have $x_1=x_2=...=x_n=1,x_{n+1}=0$ and $$x_1x_2...x_n-2\sum_{i=1}^{n}x_ix_{n+1}+(2n-1)x_{n+1}=1>0.$$
$Case2$:  $x_1x_2...x_n-x_{n+1}=0$, then we have $x_{n+1}(n-x_1-x_2...-x_n)=0$ and $$x_1x_2...x_n-2\sum_{i=1}^{n}x_ix_{n+1}+(2n-1)x_{n+1}=x_1x_2...x_n-x_{n+1}+2x_{n+1}(n-x_1-x_2...-x_n)=0.$$
$Case3$:  $x_1x_2...x_n-x_{n+1}=-1$, then we have $x_1x_2...x_n=0,x_{n+1}=1$, it implies $n-x_1-x_2...-x_n \ge 1$ and $$x_1x_2...x_n-x_{n+1}+2x_{n+1}(n-x_1-x_2...-x_n)\ge -1+2x_{n+1}=1.$$ 
In summary, the conclusion holds.$\hfill\blacksquare$

By \textbf{Observation 2.}, we can obtain the following reduction algorithm.
\begin{algorithm}[H]
  \caption{ReduceMin2($f$)}  
\hspace*{0.02in} {\bf Input:} 
A pseudo-Boolean function $f$ given by its multi-linear polynomial form.\\
\hspace*{0.02in} {\bf Output:} 
A quadratic pseudo-Boolean function $g$.
 \begin{algorithmic}[1] 
\State {\bf Initialize:} Set $M=1+2\sum_{S\subseteq [n]}|c_S|$, $m=n$, and $f^n=f$.
\While{there exists a subset $S^*\subseteq[n]$ for which $|S^*|>2$ and $c_{S^*}\neq 0$}
\State 1. Choose a $S'\subset S^*$ with size $\lceil|S^*|/2 \rceil $ and update
\State$c_{S'}=c_{S'}+M$, set
\State for all $s\in S'$ define
\State$c_{ \{s, m+1\}}= -2M$ and
\State$c_{\{m+1\}}= (2\lceil|S^*|/2 \rceil-1)M$, and
\State for all subset $S$, $S'\subseteq S$ with $c_S \neq 0$ define 
\State  $c_{(S\backslash S')\cup \{m+1\}}=c_S$ and set $c_S=0$.
\State 2. Define $f^{m+1}(x_1,...,x_{m+1})=\sum_{S\subseteq [n]}c_S\prod_{k\in S}{x_k}$, and set $m=m+1$. 
\EndWhile
\State Output: $g=f^m$
    \label{code:recentEnd}  
  \end{algorithmic}  
\end{algorithm}

Compared with \textbf{Algorithm1}, \textbf{Algorithm2} can reduce the degree of the function more quickly when the degree of the function is high. For example, $f=x_1x_2x_3x_4x_5x_6x_7$, choose $S'=\{1,2,3,4\}$, replace $x_1x_2x_3x_4$ with $x_8$ and introduce a penalty term $x_1x_2x_3x_4-2(x_1x_8+x_2x_8+x_3x_8+x_4x_8)+7x_8$, then we get
$$g=x_5x_6x_7x_8+x_1x_2x_3x_4-2(x_1x_8+x_2x_8+x_3x_8+x_4x_8)+7x_8.$$
$g$ has the same minimum with $f$ and its degree is only 4. It can be seen that we can halve the degree of a high-order term by one substitution. On the contrary, \textbf{Algorithm 1} requires multiple substitutions to do the same thing. Unfortunately, the problem we discuss in this paper requires us to reduce the degree of the functions to 2, in which case Algorithm 2 has no advantage over Algorithm 1 (and in some cases even worse). But if our goal is not to reduce a pseudo-Boolean function to quadratic but simply to reduce the degree, \textbf{Algorithm 2} has a huge advantage over \textbf{Algorithm 1} in the number of auxiliary variables. 

The optimization of a quadratic pseudo-Boolean function, also known as QUBO(Quadratic Unconstrained Binary Optimization), is a significant problem which can be solved by quantum annealing. \textbf{Algorithm 1} provides an effective approach to transform the optimization of a pseudo-Boolean function to QUBO. However, the coefficients of polynomial output in \textbf{Algorithm 1} sometimes are too large in practice, far beyond the processing scope of D-Wave computers. We first optimize algorithm 1 for some special cases.

Solving a system of pseudo-Boolean equations is a fundamental problem in the fields of mathematics and computing, and has abroad application in cryptography, machine learning and artificial intelligence. Given a system of pseudo-Boolean equations
\begin{equation}
    \begin{cases}
        f_1 = k_1 \\
        f_2 = k_2 \\
        ...\\
        f_m = k_m
     \end{cases}
\end{equation}
It can be written as $\sum_{i=1}^m{(f_i-k_i)^2}=0$, which may be also viewed as a problem of minimizing a pseudo-Boolean function due to the nonnegative properties of $f$. For this special case, there is another algorithm to ruduce it to QUBO which can greatly reduce the range of coefficients compared with \textbf{Algorithm 1}.
\begin{algorithm}[H]
  \caption{ReduceMin3($f$)}  
\hspace*{0.02in} {\bf Input:} 
A pseudo-Boolean function $f$ given by $f=\sum_{i=1}^k{f_i^2}$, \\ 
 \hspace*{0.5cm} where $f_i(x_1,...,x_n)=\sum_{S\subseteq [n]}c_S^i\prod_{j\in S}x_j$, and $\exists \textbf{x}\in F_2^n$ s.t. $f_i(\textbf{x})=0$\\.
\hspace*{0.02in} {\bf Output:} 
A quadratic pseudo-Boolean function $g$.
  \begin{algorithmic}[1]  
\State {\bf Initialize:} $m=n$, $f^n=f$, and $p=0$
\While{there exists a subset $S^*\subseteq[m]$ for which $|S^*|\geq 2$ and $c_{S^*}^t\neq 0$ for some $t \leq k$}
\State Choose two elements $i$ and $j$ from $S^*$ and update
\For{$t=1$ to $k$}
\For{all subset $S \supseteq \{i,j\}$ with $c_S^t \neq 0$}
\State  $c_{(S\backslash \{i,j\})\cup \{m+1\}}^t=c_S^t$ and set $c_S^t=0$.
\EndFor
\State $f_t^{m+1}(x_1,...,x_{m+1})=\sum_{S\subseteq [n]}c_S^t\prod_{k\in S}{x_k}$
\EndFor
\State Set $p=p+x_ix_j-2x_ix_{m+1}-2x_jx_{m+1}+3x_{m+1}$ and $m=m+1$.
\EndWhile
\State Output g=$\sum_{i=1}^k{(f_i^{m})^2}+p$
    \label{code:recentEnd}  
  \end{algorithmic}  
\end{algorithm}

In algorithm ReduceMin2($f$), we replace each ocurrence of $x_ix_j$ in all $f_t$ by $x_{m+1}$, and add the penalty term $x_ix_j-2x_ix_{m+1}-2x_jx_{m+1}+3x_{m+1}$ to the objective function. Then we have 
$$min_{\textbf{y} \in F_2^m} g(\textbf{y})=min _{\textbf{x} \in F_2^n} f(\textbf{x}) =0.$$
If and only if all penalty terms are $0$, $g$ goes to a minimum. So we can find the minimum point of $f$ by minimizing $g$.
\subsection{Quadratic optimization}
We can express the QUBO model by the following optimization problem:
$$min_{x\in \{0,1\}^n}x^TQx,$$
where Q is an $N\times N$ upper-diagonal matrix of real weights, and $x$ is a vector of binary variables. Moreover, diagonal terms $Q_{i,i}$ are linear coefficients, and the nonzero off-diagonal terms are quadratic coefficients $Q_{i,j}$.

QUBO problem may be also viewed as a problem of minimizing the function
$$f(x)=\sum_i{Q_{i,i}x_i}+\sum_{i<j}{Q_{i,j}x_ix_j}.$$

Let $x_i=\frac{1-s_i}{2}$, $s_i\in \{1,-1\}$, then $f(x)$ can be replaceed by 
\[ g(s)=\sum_i{Q'_{i,i}s_i}+\sum_{i<j}{Q'_{i,j}s_is_j}.\]
At present, small-scale QUBO problems can be solved directly by quantum annealer. Q corresponds to the ising model as follows
$$h^T=(Q'_{1,1}\ \ Q'_{2,2} \ \ ... \ \ Q'_{n,n})$$

\[
J =
\begin{pmatrix}
& & Q'_{1,2} & Q'_{1,3}& \cdots&Q'_{1,n} & \\
 & &  & Q'_{2,3} & \cdots& Q'_{2,n} &\\
& &  &  & \ddots &\vdots\\
& & &  &  & Q'_{n-1,n} &\\
& & & & & & \\
\end{pmatrix}
\]

But limited by hardware conditions, the problem becomes difficult when the number of variables $N$ is large. D-wave 2000Q, the largest quantum annealer at present with 2000 qubits, can only handle about 100 variables. 

\textbf{Theorem 1} and \textbf{Theorem 2} shows the connection between a posiform and a graph with nonnegative weights. We will perform variable reduction for the QUBO problem with this interesting connection.

First of all, we transform the minimization of a quadratic pseudo-Boolean function $f$ into the maximization of a posiform. Observe that 
$$-x_i=-1+\overline{x_i}\ \ \  -x_ix_j=-1+\overline{x_j}+\overline{x_i}x_j,$$
then one can take the negative of $f$ and complete the transformation by the observation.

For example, 
$$g=58x_1+50x_2+12x_3-80x_4+25x_1x_2-6x_1x_3-64x_1x_4+2x_2x_3-64x_2x_4+16x_3x_4+149.$$ which is the energy function of factoring $N=15=5 \times 3$, we can transform the minimization of $g$ into the maximization of 
$$g'=58\overline{x_1}+75\overline{x_2}+14\overline{x_3}+64x_4+25\overline{x_1}x_2+6x_1x_3+64x_1x_4+2\overline{x_2}x_3+64x_2x_4+16\overline{x_3}x_4.$$

By \textbf{Theorem 1}, $max_{x\in {F_2^n}}g'(x)=\alpha_{a}(G_{g'})$ 
\begin{figure}[h]
\caption{The conflict graph of $g'$}
$$\begin{tikzpicture} 
\draw[color=black,thick] (0,0)--(2,0);
\draw[color=black,thick] (2,0)--(2,4);
\draw[color=black,thick] (2,0)--(0,2);
\draw[color=black,thick] (0,2)--(2,4);
\draw[color=black,thick] (2,0)--(4,2);
\draw[color=black,thick] (2,4)--(4,2);
\draw[color=black,thick] (6,4)--(4,2);
\draw[color=black,thick] (6,2)--(4,2);
\draw[color=black,thick] (6,0)--(4,2);
\node[left] at (4,2) {$\overline{x_1}x_2$};
\node[above] at (2,4) {$x_1x_3$};
\node[left] at (0,2) {$\overline{x_3}x_4$};
\node[below] at (2,0) {$\overline{x_2}x_3$};
\node[above] at (6,4) {$\overline{x_2}$};
\node[left] at (6,0) {$x_1$};
\node[left] at (0,0) {$x_2x_4$};
\node[left] at (4,4) {$x_4$};
\node[above] at (6,2) {$x_1x_4$};
\node[left] at (2,2) {$\overline{x_3}$};
\filldraw [black] (0,0) circle (1.5pt);
\filldraw [black] (2,0) circle (1.5pt);
\filldraw [black] (2,4) circle (1.5pt);
\filldraw [black] (2,2) circle (1.5pt);
\filldraw [black] (0,2) circle (1.5pt);
\filldraw [black] (4,2) circle (1.5pt);
\filldraw [black] (4,4) circle (1.5pt);
\filldraw [black] (6,4) circle (1.5pt);
\filldraw [black] (6,2) circle (1.5pt);
\filldraw [black] (6,0) circle (1.5pt);
\end{tikzpicture}$$
\end{figure}
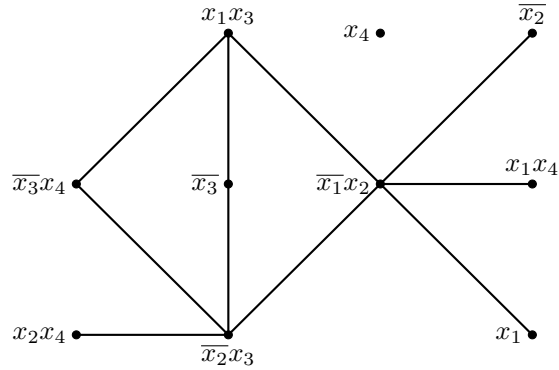

Then we will show how to subtract variables using some structures of the graph.
\\
\\
\textbf{Lemma 1.}  \textit{Given a weighted graph $G_\phi$, if there is an isolated point  $x\in G_\phi$, then for  any  weighted maximum independent vertex set $S\ of\ G_\phi$, $x \in S$ always holds, furthermore, when $\phi$ reaches a maximum, x is always equal to 1.}
\\
\\
\textbf{Proof.}  If $S\subseteq G_\phi$ is a maximum independent vertex set, then the terms in $S$ have no conflicting literals, and thus all of them can be made equal to 1, at the moment, all the other terms are equal to 0 because of the maximality of $S$.

By \textbf{Theorem1}, when $\phi$ reaches a maximum, the terms of $\phi$ which are equal to 1 form a maximum independent vertex set. Since $x\in S$ holds for every maximum independent vetex set $S$, we know that $x$ is equal to 1. $\hfill\blacksquare$

Then we have $x_4=1$ when $g'$ reaches a maximum, $g'$ can be reduced to 
$$g''=6x_1+11\overline{x_2}+30\overline{x_3}+25\overline{x_1}x_2+6x_1x_3+2\overline{x_2}x_3.$$ 

Similarly, $g''$ corresponds to a conflict graph $G_{g''}$.
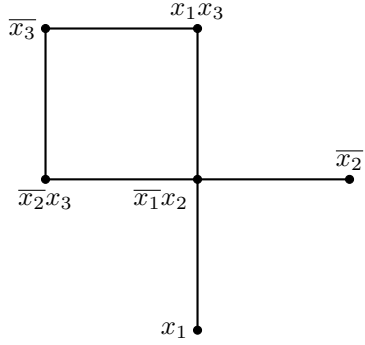
\begin{figure}[h]
\caption{The conflict graph of $g''$}
$$\begin{tikzpicture} 
\draw[color=black,thick] (2,0)--(2,4);
\draw[color=black,thick] (0,2)--(4,2);
\draw[color=black,thick] (0,2)--(0,4);
\draw[color=black,thick] (2,4)--(0,4);
\node[below left] at (2,2) {$\overline{x_1}x_2$};
\node[above] at (2,4) {$x_1x_3$};
\node[below] at (0,2) {$\overline{x_2}x_3$};
\node[above] at (4,2) {$\overline{x_2}$};
\node[left] at (2,0) {$x_1$};
\node[left] at (0,4) {$\overline{x_3}$};
\filldraw [black] (2,0) circle (1.5pt);
\filldraw [black] (2,4) circle (1.5pt);
\filldraw [black] (2,2) circle (1.5pt);
\filldraw [black] (0,2) circle (1.5pt);
\filldraw [black] (0,4) circle (1.5pt);
\filldraw [black] (4,2) circle (1.5pt);
\end{tikzpicture}$$
\end{figure}
\\
\\
\textbf{Lemma 2.}  \textit{Given a weighted graph $G_\phi$, if there are two vertices with the same adjacent points, remove one of the two vertices and add its weight to another, we get a new weighted graph $G_{\phi'}$, and a new posiform $\phi'$. Then $max_{x\in {F_2^n}}\phi(x)=max_{x\in {F_2^n}}\phi'(x)=\alpha_{a}(G_{\phi})=\alpha_{a}(G_{\phi'})$ holds.}
\\
\\
\textbf{Proof.} Without loss of generality, let $x,y\in G_{\phi}$ be the two vertices with the same adjacent points, $z$ be the new vertex in $G_{\phi'}$. For  any  weighted maximum independent vertex set $S\ of\ G_\phi$, either $x,y \in S$ or $x,y \notin S$ holds. Otherwise, if $x\in S, y \notin S$, $x,y$ share the same adjacent points, then $S \cup \{ y\}$ is also an independent vertex set, contradicting to the maximality of $S$. Thus, when $\phi$ reaches a maximum, $x=y$ holds. It follows that $max_{x\in {F_2^n}}\phi(x) \le max_{x\in {F_2^n}}\phi'(x)$. 

On the other hand, for any weighted maximum independent vertex set $S'\ of\ G_{\phi'}$, it can be mapped to an independent vertex set $S\ of\ G_{\phi}$ by split $z$ into $x$ and $y$, it follows that $\alpha_{a}(G_{\phi'})\le \alpha_{a}(G_{\phi})$. By \textbf{Theorem1}, the lemma holds.$\hfill\blacksquare$

Return to $G_{g''}$, $\overline{x_2}x_3$ and $x_1x_3$ have the same adjacent points. The same thing happens to $\overline{x_2}$ and $x_1$. Then $g''$ can be reduced to 
$$g^{(3)}=17x_1+25\overline{x_1}x_2+8x_1x_3+30\overline{x_3}.$$
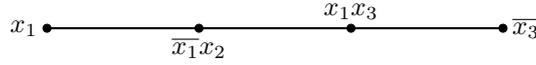
\begin{figure}[h]
\caption{The conflict graph of $g^{(3)}$}
$$\begin{tikzpicture} 
\draw[color=black,thick] (0,0)--(6,0);
\node[below] at (2,0) {$\overline{x_1}x_2$};
\node[above] at (4,0) {$x_1x_3$};
\node[left] at (0,0) {$x_1$};
\node[right] at (6,0) {$\overline{x_3}$};
\filldraw [black] (2,0) circle (1.5pt);
\filldraw [black] (0,0) circle (1.5pt);
\filldraw [black] (4,0) circle (1.5pt);
\filldraw [black] (6,0) circle (1.5pt);
\end{tikzpicture}$$
\end{figure}
\\
\\
\textbf{Corollary 1.} \textit{If two posiforms $\phi$ and $\phi'$ have the same conflict graph, then they have the same maximum, further more, one can get the maximum point of $\phi$ by solve $\phi'$.}

By \textbf{Corollary 1.}, we can get the maximum point of $g^{(3)}$ by solve 
$$g^{(4)}=17y_1+25\overline{y_1}y_2+8\overline{y_2}+30{y_2}.$$
The number of variables is reduced from $4$ to $2$.

In fact, for a given graph $G$, it always can be transformed to a new graph $G'$ by the methods in \cite{ebenegger1984pseudo} such that
$$\alpha(G)=\alpha(G')+1$$
holds. Repeating this transformation, one can arrive to a trivial graph. Although the sequence of graphs produced in this way may have an exponential growth in size. There are also some graphs with the specific structures can be reduced to trivial cases in polynomial time. The quadratic optimization problems corresponding to these graphs can , in principle, be effectively solved in the quantum environment after optimization by our method. Here we only focus on some practical problems and some specific structures to show the superiority of our method.
\section{Applications in cryptography}
\subsection{Integers Factorization}
Given a large number $N=p\times q$, where $p,q$ are unkown prime numbers, finding $p,q$ is usually regarded as a difficult problem. Without loss of generality, we take $p=(p_{n-1}p_{n-2}...p_{1}1)_2$, $q=(q_{m-1}q_{m-2}...q_{1}1)_2$, where $p_i,q_j$ are binary numbers. Then $p=\sum_{i=1}^{n-1}2^ip_i+1$, $q=\sum_{j=1}^{m-1}2^jq_j+1$. The direct method to map factoring $N$ to a QUBO problem is to define a function $f(p_1,...,p_{n-1},q_1,...,q_{m-1})=(N-pq)^2$. This function is quartic, we can reduce it to a quadratic pseudo-Boolean function by \textbf{Algorithm 2}, we need $\begin{pmatrix} n-1\\ 2 \end{pmatrix}+\begin{pmatrix} m-1\\ 2 \end{pmatrix}$ auxiliary variables. If $m=n$, the number of auxiliary variables is $n\times (n-1)$, the total number of binary variables is $2(n-1)+n(n-1)=O(n^2)=O(log^2(N))$. The range of coefficient values is about $[-N^2,N^2]$, we can see the number of variables is acceptable, but the coefficient values are too large. Jiang et al. proposed a method using multiplication table to reduce coefficient values, they substitute a system of equations for $f$ and introduce some other auxiliary variables(carries). We follow this method and built the multiplication table for $4137131=2029\times 2039$.
\begin{figure}[h]
\caption{The multiplication table for $4137131=2029\times 2039$}
\includegraphics[width=1\linewidth]{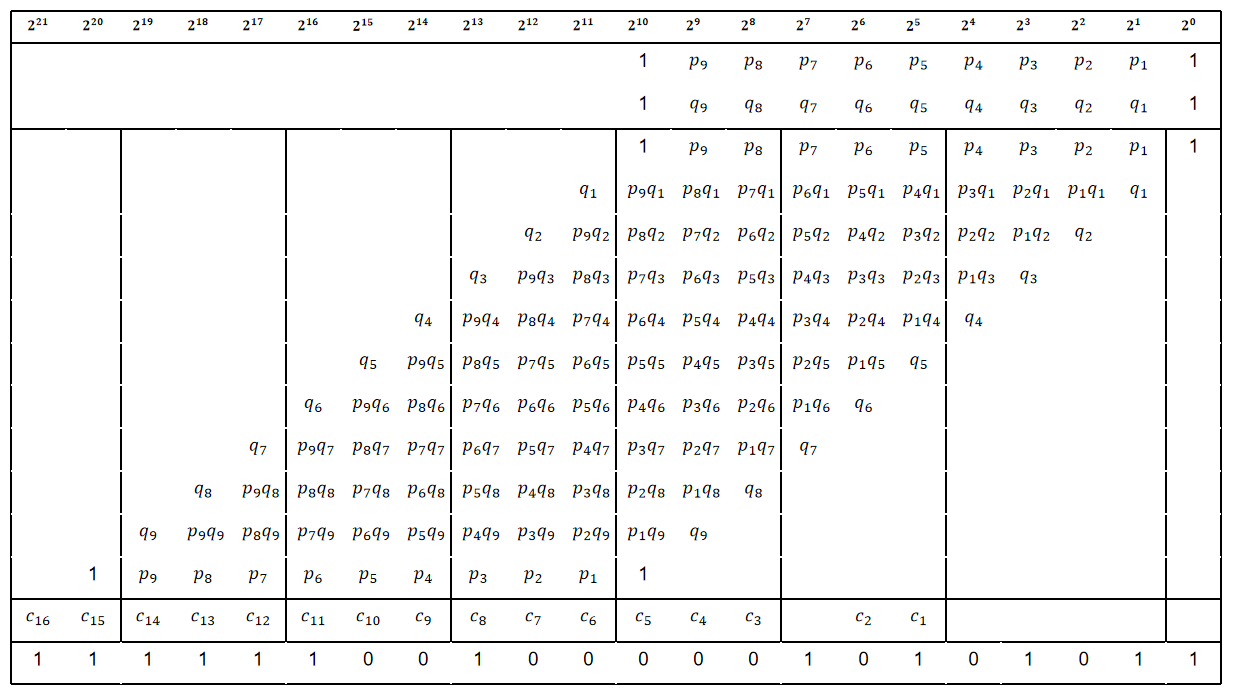}
\end{figure}

With this table, we can get seven functions $f_1$,$f_2$,...,$f_7$, for example, 
$$f_1=p_1+q_1+2(p_2+p_1q_1+q_2)+4(p_3+p_2q_1+p_1q_2+q_3)+8(p_4+p_3q_1+p_2q_2+p_1q_3+q_4)-16c_1-32c_2-5.$$
and construct a function with the form $f=\sum_{i=1}^7{f_i^2}$. By \textbf{Algorithm 3}, we can replace all the quadratic terms which appear in the table by auxiliary variables. These substitutions don't change the range of coefficient values, because the maximum and minimum coefficient are connected with $c_i$ which are not changed in \textbf{Algorithm 3}. The coefficient range of the reduced function is $[-1024,1024]$.

The number of variables is $9\times 2+9\times 9+16=115$ (every quadratic term needs an auxiliary variable). We can optimize this number to 93 with our method. For example, one can check that $c_{16}$ is an isolated point in the conflict graph of $f$, then it must be equal to 1 when $f$ reaches a maximum. We compare the original methods with our method, and the result is shown in Table 1.
\begin{table*}[h]
\caption{Comparison of results for the proposed and existing algorithms}
\centering
\begin{tabular}{|c|c|c|c|} 
\hline
               & integer $N$         & variables & range of coefficients  \\ 
\hline
Jiang's method & $376289=659\times 571$    & 95        & {[}-4848,2500]         \\ 
\hline
Wang's method  & $1028171=1009\times 1009$ & $>$89        & {[}-3005,5032]         \\ 
\hline
Our method     & $4137131=2029\times 2039$ & 93        & {[}-1024,1024]         \\
\hline
\end{tabular}
\end{table*}
\subsection{Algebraic attacks on block ciphers}
Algebraic attacks exploit the internal algebraic structure of ciphers. The general idea is obtaining a representation of
the cipher as a system of equations and trying to solve it to recover the secret key bits. In theory, most modern (block and stream) ciphers
can be described by a system of multivariate polynomials over a finite field. As we mentioned earlier, any system of pseudo-Boolean equations can be transformed into QUBO by \textbf{Algorithm 3}. So all we should do is transforming a system of Boolean equations into a system of pseudo-Boolean equations.
\subsubsection{From Boolean equations to pseudo-Boolean equations}
Given a Boolean equation
$$f(x_1,...,x_n)=\bigoplus_{S\subseteq [n]}c_S\prod_{j\in S}x_j = 0,$$
where $c_S\in F_2$, define $M_f:=\sum_{S\subseteq [n]}c_S$, $t=\lceil logM_f \rceil$, one can transform the Boolean equation into a pseudo-Boolean equation
$$\sum_{S\subseteq [n]}c_S\prod_{j\in S}x_j = \sum_{i=1}^t{2^iy_i}$$
with  $\lceil logM_f \rceil$ auxiliary variables $y_i$.
Let $f'(x_1,...,x_n)=\sum_{S\subseteq [n]}c_S\prod_{j\in S}x_j-\sum_{i=1}^t{2^iy_i}$, $f''(x_1,...,x_n)=\sum_{S\subseteq [n]}c_S\prod_{j\in S}x_j$, both $f'$ and $f''$ are pseudo-Boolean functions. Then for a system of Boolean equations $S$
\begin{equation}
    \begin{cases}
        f_1(x_1,...,x_n) = 0 \\
        f_2 (x_1,...,x_n)= 0 \\
        ...\\
        f_m (x_1,...,x_n)= 0.
     \end{cases}
\end{equation}
Substitute $f'_i$($f''_i$) for $f_i$ in $S$ to get a new system of pseudo-Boolean equations $S'$($S''$). And transform $S''$ into a quadratic pseudo-Boolean function $g''$ by \textbf{Algorithm 3}, denote the number of variables of $g''$ as $N$. The following lemma holds.
\\
\\
\textbf{Lemma 3.} $S'$ has the same solutions with $S$, and one can transform $S'$ into a quadratic pseudo-Boolean function $g'$ with at most $N+\sum_{i=1}^m{\lceil logM_{f_i} \rceil} $ variables, and the range of coefficients of $g'$ is $[-M^2,M^2]$, where $M=max\{M_{f_i}\}$.
\\
\\
\textbf{Proof.} For a solution $(x_1,...,x_n,y_1,...,y_m)$ that satisfies the equation $f'=\sum_{S\subseteq [n]}c_S\prod_{j\in S}x_j-\sum_{i=1}^t{2^iy_i}=0$, it is easy to check $\bigoplus_{S\subseteq [n]}c_S\prod_{j\in S}x_j = 0$. On the other hand, if $(x_1,...,x_n)$ satisfies the equation $\bigoplus_{S\subseteq [n]}c_S\prod_{j\in S}x_j = 0$, then $\sum_{S\subseteq [n]}c_S\prod_{j\in S}x_j\equiv0 \ (mod\ 2)$, since $\sum_{S\subseteq [n]}c_S\prod_{j\in S}x_j < \sum_{S\subseteq [n]}c_S=M_f$, there exist $y_i$ such that $\sum_{S\subseteq [n]}c_S\prod_{j\in S}x_j = \sum_{i=1}^t{2^iy_i}$.

It is noticed that $y_i$ doesn't need auxiliary variables to reduce its degree in \textbf{Algorithm 3}, the number of variables of $g'$ is only increased by $\sum_{i=1}^m{\lceil logM_{f_i} \rceil} $ compared to the system $S$.

As for the coefficients, one can easy check that the maximum and minimum coefficient are connected with $y_i$ which are not changed in \textbf{Algorithm 3}, so the range is $[-2^{2\lceil logM_f \rceil},2^{2\lceil logM_f \rceil}]\subset [-M^2,M^2]$. $\hfill\blacksquare$
\subsubsection{Algebraic attacks on AES-128}
The original AES is a 128-bit block cipher designed by Daemen and Rijmen\cite{daemen2002design}. It is based on Substitution-Permutation Network(SPN) and each round transformation consists of the four operations SubBytes(SB), ShiftRows(SR), MixColumns(MC), and AddRoundKey(AK). SubBytes is the non-linear operation that applies a 8-bit S-box to each cell. ShiftRows is the linear operation that rotates the i-th row by i cells to the left. MixColumns is the linear operation that multiplies an $r\times r$ matrix over $GF(2^c)$ to each column vector. AddRoundKey is the operation thatadds a round key to the state\cite{hosoyamada2020finding}. The round function of AES can be described as
$$MC\circ SR\circ SB\circ AK.$$

In this section, we follow Burek's method and use the following 13 quadratic Boolean equations to represent the S-box of AES.

\begin{math}
\begin{aligned}
eq_{0}: & x_{0} y_{2}+x_{0} y_{4}+x_{2} y_{0}+x_{2} y_{1}+x_{3} y_{1}+x_{3} y_{2}+x_{3} y_{7}+x_{4} y_{3}+ x_{4} y_{4}+x_{5} y_{1}+x_{5} y_{2}+\\
&x_{5}y_{3}+x_{5} y_{7}+x_{6} y_{1}+x_{6} y_{2}+x_{6} y_{3}+x_{0}+x_{2}+x_{4}+x_{7}+y_{0}+y_{5}=0, \\
eq_{1}: & x_{0} y_{3}+x_{0} y_{5}+x_{2} y_{0}+x_{2} y_{1}+x_{2} y_{5}+x_{3} y_{0}+x_{3} y_{3}+x_{3} y_{7}+x_{4} y_{0}+x_{4} y_{3}+x_{4} y_{4}+\\
&x_{5} y_{0}+x_{6} y_{3}+x_{6} y_{4}+x_{6} y_{5}+x_{7} y_{1}+x_{7} y_{5}+x_{7} y_{7}+x_{2}+x_{4}+y_{0}=0 \\
eq_{2}: & x_{0} y_{0}+x_{0} y_{2}+x_{0} y_{4}+x_{0} y_{6}+x_{1} y_{5}+x_{1} y_{7}+x_{2} y_{0}+x_{3} y_{1}+x_{3} y_{2}+x_{4} y_{0}+x_{4} y_{1}+\\
&x_{4} y_{7}+x_{5} y_{0}+x_{5} y_{3}+x_{6} y_{4}+x_{7} y_{4}+x_{5}+y_{1}+y_{7}=0, \\
eq_{3}: & x_{0} y_{5}+x_{1} y_{1}+x_{3} y_{0}+x_{4} y_{0}+x_{4} y_{1}+x_{4} y_{3}+x_{4} y_{6}+x_{5} y_{6}+x_{6} y_{1}+x_{6} y_{2}+x_{6} y_{3}+\\
&x_{7} y_{2}+x_{7} y_{4}+x_{1}+x_{4}+x_{5}+x_{6}+x_{7}+y_{4}=0,\\
\end{aligned}
\end{math}

\begin{math}
\begin{aligned}
e q_{4}: & x_{0} y_{0}+x_{0} y_{2}+x_{0} y_{4}+x_{0} y_{5}+x_{3} y_{3}+x_{3} y_{4}+x_{4} y_{4}+x_{5} y_{0}+x_{5} y_{1}+x_{6} y_{1}+x_{6} y_{5}+\\
&x_{7} y_{2}+x_{7} y_{5}+x_{0}+x_{1}+x_{2}+x_{5}+x_{6}+x_{7}+y_{5}+y_{6}+y_{7}=0, \\
e q_{5}: & x_{0} y_{0}+x_{0} y_{3}+x_{0} y_{5}+x_{2} y_{0}+x_{2} y_{2}+x_{2} y_{4}+x_{2} y_{5}+x_{3} y_{2}+x_{3} y_{3}+x_{3} y_{4}+x_{3} y_{5}+\\
&x_{4} y_{3}+x_{5} y_{0}+x_{5} y_{4}+x_{5} y_{6}+x_{6} y_{0}+ x_{6} y_{1}+x_{6} y_{2}+x_{0}+y_{1}+y_{2}=0, \\
e q_{6}: & x_{0} y_{3}+x_{0} y_{5}+x_{0} y_{7}+x_{2} y_{4}+x_{2} y_{6}+x_{2} y_{7}+x_{3} y_{2}+x_{3} y_{6}+ x_{4} y_{4}+x_{5} y_{3}+x_{5} y_{7}+\\
&x_{6} y_{4}+x_{6} y_{5}+x_{7} y_{1}+x_{7} y_{7}+x_{0}+ x_{4}+x_{5}+y_{0}+y_{3}=0, \\
e q_{7}: & x_{0} y_{4}+x_{0} y_{7}+x_{1} y_{5}+x_{2} y_{1}+x_{2} y_{6}+x_{2} y_{7}+x_{3} y_{1}+x_{3} y_{4}+x_{4} y_{0}+x_{4} y_{4}+x_{4} y_{5}+\\
&x_{4} y_{7}+x_{5} y_{4}+x_{5} y_{5}+x_{6} y_{4}+x_{7} y_{2}+x_{7} y_{7}+x_{3}+x_{5}+x_{6}+y_{0}+y_{5}=0, \\
e q_{8}: & x_{0} y_{0}+x_{0} y_{2}+x_{0} y_{7}+x_{2} y_{0}+x_{2} y_{3}+x_{2} y_{4}+x_{3} y_{4}+x_{3} y_{6}+x_{4} y_{4}+x_{5} y_{1}+x_{6} y_{0}+\\
&x_{6} y_{1}+x_{6} y_{2}+x_{6} y_{4}+x_{7} y_{2}+x_{7} y_{3}+ x_{6}+y_{1}+y_{2}+y_{5}+y_{7}+1=0,\\
e q_{9}: &x_{0} y_{0}+x_{0} y_{4}+x_{0} y_{6}+x_{0} y_{7}+x_{2} y_{7}+x_{3} y_{3}+x_{4} y_{0}+x_{4} y_{1}+x_{4} y_{3}+x_{4} y_{6}+x_{5} y_{4}+\\
&x_{5} y_{6}+x_{5} y_{7}+x_{6} y_{1}+x_{6} y_{3}+x_{6} y_{4}+x_{7} y_{1}+x_{7} y_{3}+x_{2}+y_{3}+y_{5}=0 ,\\
e q_{10}: &x_{0} y_{2}+x_{0} y_{7}+x_{1} y_{5}+x_{1} y_{7}+x_{2} y_{0}+x_{2} y_{2}+x_{3} y_{4}+x_{3} y_{7}+x_{4} y_{2}+x_{4} y_{3}+x_{5} y_{4}+\\
&x_{7} y_{2}+x_{7} y_{5}+x_{7} y_{7}+x_{1}+x_{7}+y_{3}+y_{5}+y_{7}+1=0 \text {, }\\
e q_{11}: &x_{2} y_{3}+x_{2} y_{4}+x_{3} y_{4}+x_{4} y_{0}+x_{4} y_{1}+x_{5} y_{1}+x_{5} y_{5}+x_{6} y_{2}+x_{6} y_{5}+x_{7} y_{1}+x_{0}+\\
&x_{4}+x_{5}+x_{6}+y_{4}+y_{5}+y_{6}+1=0 \text {, }\\
e q_{12}: &x_{0} y_{4}+x_{0} y_{6}+x_{1} y_{7}+x_{2} y_{1}+x_{2} y_{4}+x_{2} y_{5}+x_{2} y_{6}+x_{3} y_{0}+x_{3} y_{2}+x_{3} y_{4}+x_{3} y_{6}+\\
&x_{4} y_{1}+x_{4} y_{3}+x_{5} y_{0}+x_{5} y_{1}+x_{5} y_{2}+x_{6} y_{5}+x_{4}+y_{1}+y_{5}+y_{7}=0.\\
\end{aligned}
\end{math}
\\

Unlike Burek, we focus more on the optimized coefficients of the equations. By \textbf{Algorithm 3} and \textbf{Lemma 3}, we can transform the above Boolean equarions into a pseudo-Boolean function with coefficients ranging [-256,256], more precisely, [-32,256]. For example, $eq_0$ can be rewritten as a pseudo-Boolean equation $eq_0'$:
\\
\\
\begin{math}
\begin{aligned}
& x_{0} y_{2}+x_{0} y_{4}+x_{2} y_{0}+x_{2} y_{1}+x_{3} y_{1}+x_{3} y_{2}+x_{3} y_{7}+x_{4} y_{3}+ x_{4} y_{4}+x_{5} y_{1}+x_{5} y_{2}+x_{5}y_{3}+\\
&x_{5} y_{7}+x_{6} y_{1}+x_{6} y_{2}+x_{6} y_{3}+x_{0}+x_{2}+x_{4}+x_{7}+y_{0}+y_{5}-16z_1-8z_2-4z_3-2z_4=0. \\
\end{aligned}
\end{math}
\\

And $eq_0'^2$ can be reduced to a quadratic pseudo-Boolean function with the form
$$(\sum_{i=1}^{16}u_i+x_{0}+x_{2}+x_{4}+x_{7}+y_{0}+y_{5}-16z_1-8z_2-4z_3-2z_4)^2+Pen,$$
where $u_i$ is the substitution for the quadratic term in $eq_0$, $Pen$ is the sum of all the penalties. The range of cofficients of $Pen$ is [-2,3] by \textbf{Algorithm 3}. The coefficient range of the total pseudo-Boolean function is [-32,256]. The results for specific coefficients are shown in Table 2.
\begin{table*}[h]
\caption{Results of Transformation of AES-128 to the QUBO Problem}
\centering
\begin{tabular}{|c|c|} 
\hline
\makecell[c]{The coefficient range of equations in the system\\ describing the substitution box. }       & [-32,256]         \\ 
\hline
\makecell[c]{The coefficient range of equations in the system\\ describing SR and MC. } & [-8,16]         \\ 
\hline
\makecell[c]{The coefficient range of equations in the system\\describing key expansion. }  & [-32,256]         \\ 
\hline
\makecell[c]{The coefficient range of equations in the system \\describing the cipher.} & [-32,256]          \\
\hline
\end{tabular}
\end{table*}
\section{Discussion and conclusions}
In this paper, we present a new method to reduce the number of variables and coefficient range in QUBO, it can be used to reduce the number of qubits required for integer decomposition and algebraic attack on block ciphers. We have achieved the largest integer factorization in quantum computers by our own method. But the quantum algebraic attack on modern block ciphers like AES seems to be still a long way to go. Our method can be also used in other fields such like machine learning and natural language processing. While the optimization techniques used in this paper are very simple, a topic of future work is to apply much more powerful techniques to achieve better optimizations.
\bibliographystyle{alpha}
\bibliography{abbrev3,bilbio,crypto}

\end{document}